\documentclass{emulateapj}
\usepackage{times}

\begin{document}
\markboth{Cheng-Jiun Ma} {Dynamics of the core of MACSJ0717.5+3745}

\title{An X-ray/optical study of the complex dynamics of the core of the massive
  intermediate-redshift cluster MACSJ0717.5+3745} \author{Cheng-Jiun Ma, 
  Harald Ebeling, and Elizabeth Barrett} \affil{Institute for Astronomy, University of Hawaii, 2680
  Woodlawn Drive, Honolulu, HI 96822, USA}

\begin{abstract}
  Using {\it CHANDRA}, we investigate the spatial temperature distribution of
  the intracluster medium (ICM) within 700 kpc of the center of the massive
  merging cluster MACSJ0717.5+3745 at $z=0.55$. Combining the X-ray evidence
  with information about the distribution and velocities of the cluster galaxies
  near the core provides us with a snapshot of the three-dimensional geometry
  and dynamics of one of the most complex cluster studied to date. We find
  MACSJ0717.5+3745 to be an active triple merger with ICM temperatures 
  exceeding 20 keV. Although radial velocity information and X-ray/optical 
  offsets indicate that all three mergers proceed along distinctly different 
  directions, the partial alignment of the merger axes points to a common 
  origin in the large-scale filament south-east of the cluster core. Clear 
  decrements in the ICM temperature observed near two of these subclusters 
  identify the respective X-ray surface brightness peaks as remnants of cool 
  cores; the compactness and low temperature of 5.7 keV of one of these 
  features suggest that the respective merger, a high-velocity collision at 
  3,000 km s$^{-1}$, is still in its very early stages. Looking beyond the
  triple merger, we find the large-scale filament to not only provide a 
  spatial as well as temporal arrow for the interpretation of the dynamics of 
  the merger events near the cluster core; we also find tantalizing, if 
  circumstantial, evidence for direct, large-scale heating of the ICM by 
  contiguous infall of low-density gas from the filament.
 
\end{abstract}


\section{Introduction}\label{intro}

Major cluster mergers are the most energetic events in the universe, and one of
the most important phenomena for the formation of large-scale structure in the
hierarchical scenario. The enormous energy released in cluster mergers
accelerates particles to relativistic energies and can, in the most massive
systems, heat the intracluster medium (ICM) to temperatures exceeding 20 keV
\citep[see review by][]{sarazin02}. In an astrophysical context, dynamically
simple two-component mergers like 1ES0657--56 have been instrumental in obtaining
the currently best constraint on the self-interaction cross section of dark
matter \citep{markevitch04,randall08}\footnote{see also
  \citet{mahdavi07,bradac08} for results obtained for A\,520 and
  MACSJ0025.4--1222.}.  Few mergers, however, are as straightforward to
interpret as 1ES0657--56 which consists of only two well separated components,
and whose merger axis is almost perpendicular to our line of sight. While the
majority of massive mergers are thus among the most challenging systems for
studies of the dynamics and interactions of gas, galaxies, and dark matter, they
also offer the greatest opportunity to study with good statistics the full
complexity of cluster mergers.

A prime example of a complex, major cluster merger is MACSJ0717.5+3745
($z=0.545$). Discovered during the Massive Cluster Survey
\citep[MACS][]{ebel01,ebel07}, the system is one of the most X-ray luminous and
best-studied massive clusters at intermediate redshifts. The wealth of data
accumulated for it from X-ray to mm wavelengths make MACSJ0717.5+3745 a
promising target for a comprehensive study of the full range of physical
mechanisms at work during the assembly of massive clusters. At radio
frequencies, synchroton emission from relativistic electrons in MACSJ0717.5+3745
was detected in the form of both a radio halo and a very extended,
steep-spectrum radio relic \citep{edge03}. The same relativistic electrons are
expected to also produce hard X-ray photons via inverse Compton scattering of
Cosmic Microwave Background photons, as reported by \citet{petrosian06}. The
spatial distribution of the galaxy population in and around MACSJ0717.5+3745 was
studied by \citet{ebel04} who found the cluster to be connected to an extended
linear filamentary structure, consistent with the picture seen in numerical
simulations \citep[and observationally supported by the findings of][]{jeyh08}
that the most massive halos are more likely to be surrounded by pronounced
large-scale structure \citep[e.g.][]{colberg00,colberg99}. Finally, a
spectroscopic analysis by Ma et al.\ (2008) found strong evidence for star
formation in many galaxies in and around this cluster, likely caused by the
ongoing growth of this complex system. In this letter, we study variations in
the temperature of the ICM of MACSJ0717.5+3745 and their correlation with 
features in the spatial and radial-velocity distribution of the cluster galaxy distribution, 
in order to assess the dynamical state and history of the cluster core.

\section{X-ray data}\label{data-xray}

MACSJ0717.5+3745 was observed with the Chandra Advanced CCD Imaging Spectrometer
Imaging array \citep[ACIS-I;][]{acisi03} in January 2003 for a total exposure
time of 60 ks in VFaint mode (ObsID 4200). We applied standard Chandra data
reduction
procedures\footnote{$\url{http://cxc.harvard.edu/ciao/guides/acis\_data.html}$}
to reprocess the event file using CIAO 3.3.0.1 and CALDB 3.2.1. Our
spectroscopic analysis made use of a ``blank-sky'' background file\footnote{{see
    http://cxc.harvard.edu/contrib/maxim/acisbg/COOKBOOK}} to remove 
low-level soft X-ray emission covering a majority of the ACIS-I array.

\begin{figure}[t]
  \plotone{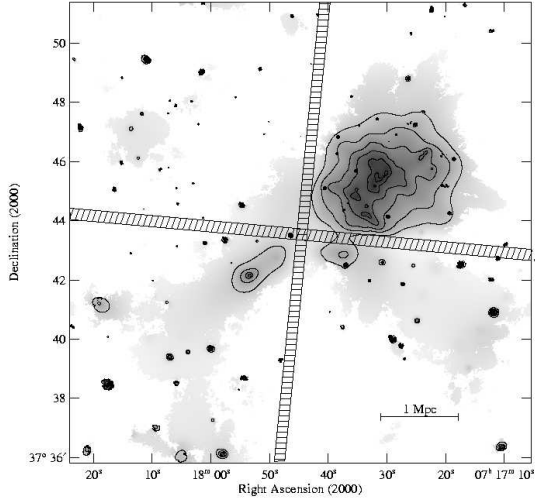}
  \figcaption[Xray image]{X-ray emission in the field of MACSJ0717.5+3745 as
    observed with Chandra's ACIS-I detector in the 0.5--7 keV band. The image
    has been weighted by the exposure map and adaptively smoothed using the
    \textit{asmooth} algorithm \citep{ebel06} requiring a minimal significance
    of $99\%$ with respect to the local background \citep[see also][]{ma08,
      ebel07}. The shaded region marks the chip gaps. Contours are
    logarithmically spaced by factors of two, starting at six times the value of
    the global X-ray background. \label{xrayim}}
\end{figure}   

An adaptively smoothed representation of the data within the full field of view
of ACIS-I is shown in Fig.~\ref{xrayim}. The disturbed X-ray morphology around
the core of the cluster, which contains at least four X-ray surface-brightness
peaks, is immediately obvious in the upper right corner. In
addition, weak but detectable X-ray emission extends to the south-eastern corner
of Fig.~\ref{xrayim}, matching the large-scale filament
discovered in the galaxy distribution by \citet{ebel04}.\footnote{This includes
  the satellite cluster 2 Mpc south-east of the main system; we measure its
  ICM temperature to be $2.7\pm0.3$ keV.}

\section{Optical data}\label{data-opt}

MACSJ0717.5+3745 was observed with the Advanced Camera for Surveys (ACS) aboard
the Hubble Space Telescope (HST) in the F555W and F814W passbands on April 2,
2004 for an effective exposure time of 4470 and 4560s, respectively, as part of
a comprehensive HST survey of the most distant MACS clusters
(GO-09722). Groundbased observations of the galaxy population of the cluster
were conducted with Keck-II/DEIMOS from 2004 to 2008 with the aim of probing
large-scale cluster dynamics and examining the effect of environment on galaxy
evolution (Ma et al.\ 2008).

\section{X-ray temperature map}\label{xray-temp}

As is apparent from its complex X-ray morphology, MACSJ0717.5+3745 is far from
virialized, and the gas temperature can be expected to vary significantly in its
core\footnote{A summary of the {\it global}\/ properties of the hot gas in
  MACSJ0717.5+3745 is given by \citet{ma08}.}. In order to derive a spatial
temperature map, we need to define suitable regions for X-ray
spectroscopy\footnote{We perform all spectral fitting with \textit{Sherpa} using
  a MEKAL plasma model \citep{mewe85}, with the absorption term fixed at the
  Galactic value of nH $= 7.11\times 10^{20}{~\rm cm}^{-2}$, the redshift of
  the cluster set to z$_{cl}=0.5446$ \citep{ma08}, and abundance Z $=0.3Z_{\sun}$.}. Ideally, these regions
should follow the temperature structure. The latter, however, is {\it a
  priori}\/ unknown, and the limited photon statistics of our data do not allow
us to extract spectra in arbitrarily small bins. We address this problem by
adopting a two-level binning method. As a first step, we subdivide the image
into regions whose shape is guided by the X-ray surface-brightness distribution,
and each of which contains enough photons to allow a crude temperature
measurement. We then merge adjacent regions with similar temperature in order to 
obtain a clearer view of temperature variations on larger scales. This procedure ensures that the final regions
represent the temperature structure fairly, and also that the
resulting measurements are statistically robust.

\subsection{Low-fidelity temperature map}\label{xray-temp-1}
The initial regions for this process are determined by
running the "accumulative binning" algorithm \textit{contbin} \citep{sanders06}
on the exposure-weighted image\footnote{In \textit{contbin}\/ the image was
  smoothed to a signal-to-noise ratio (SNR) of 10, and we set the "constraint
  parameter" $C$ \citep{sanders06} to 1.5 to prevent the bins from becoming too
  elongated. The SNR required of the resulting regions was 20, which corresponds
  to 400 net counts in regions with negligible background.}. We chose
\textit{contbin} because the algorithm's choice
of bins is physically motivated, in as much as \textit{contbin} attempts to
follow the X-ray surface brightness, which is related to the ICM temperature and
density. If the initial regions are sufficiently small, their choice will, by
the end of the merging process, not affect the final temperature map.

\begin{figure*}
  \epsscale{1.0}
    \plottwo{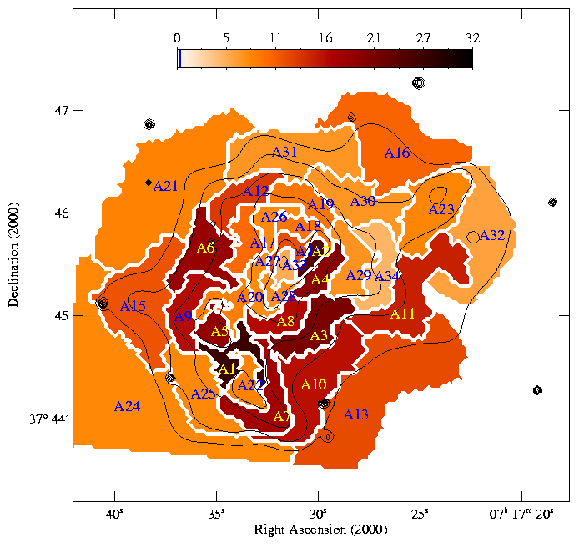} {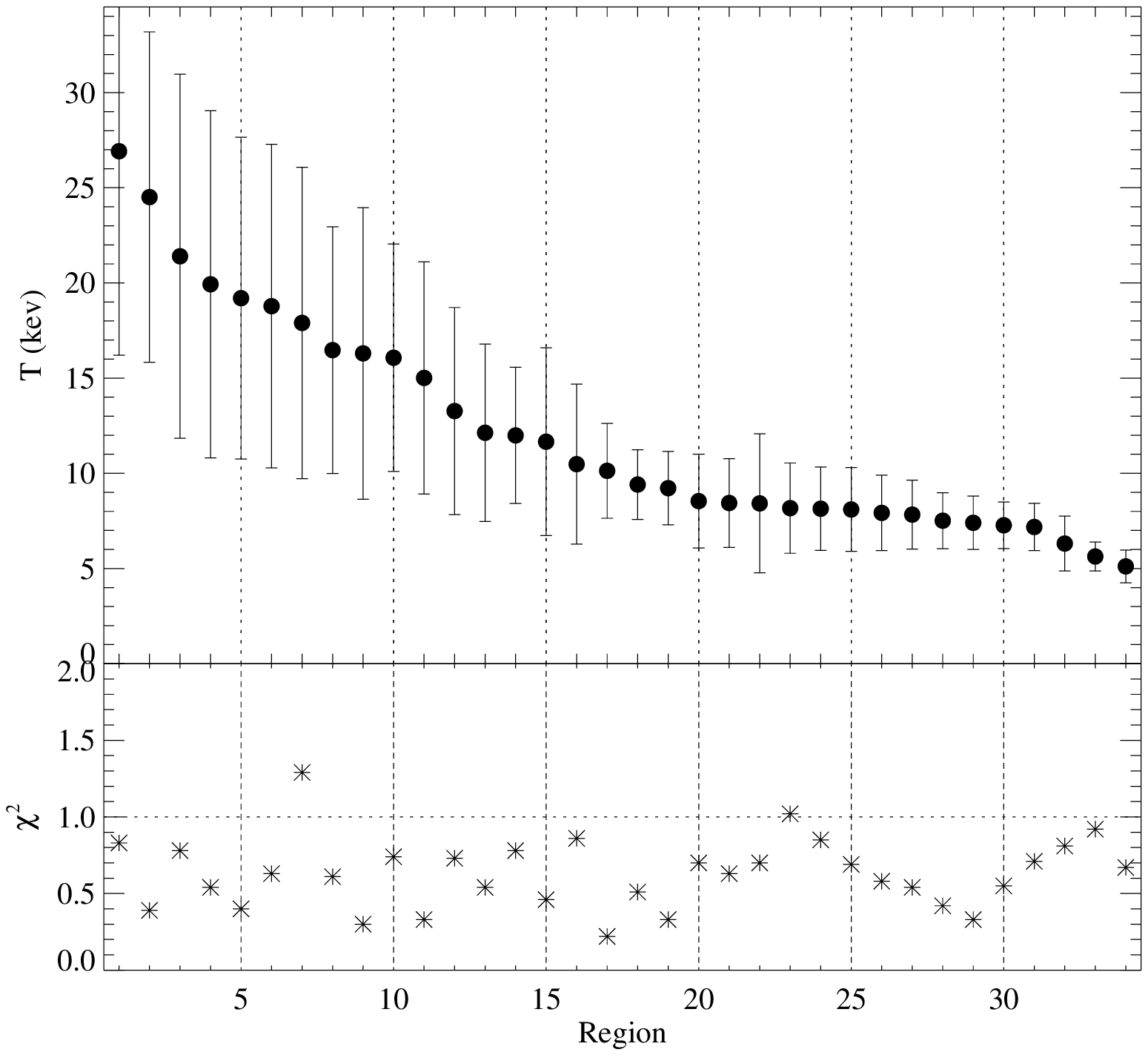}
  \vspace{-5mm}   \plottwo{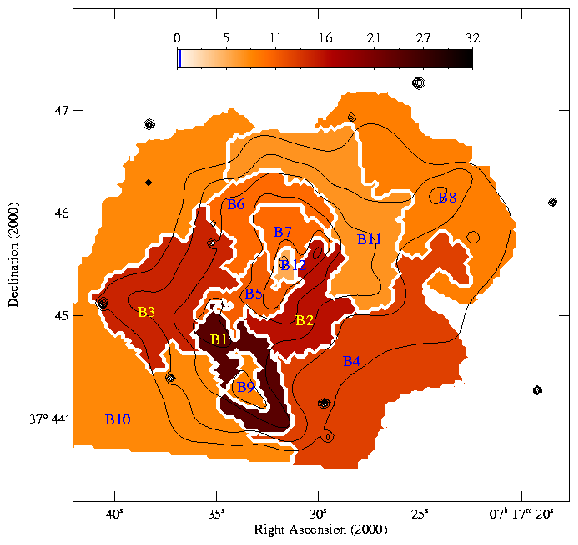} {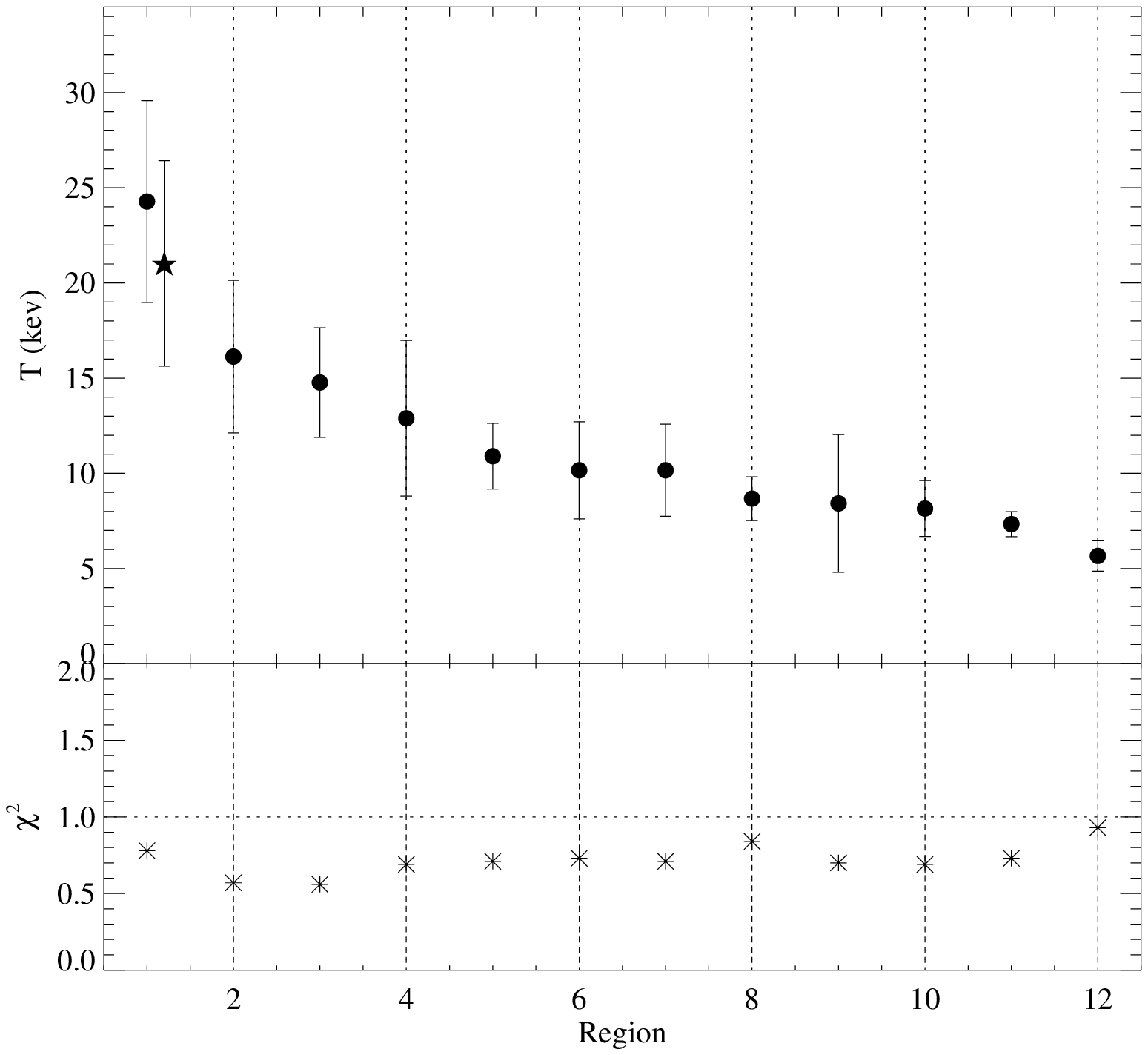}
  \figcaption[xtemp1]{Left panels: X-ray temperature maps from our two-step
    binning process. The temperature is coded as indicated by the color
    bar. Right panels: the best-fit temperature and reduced $\chi^2$ value for
    each bin. Top: results for the initial regions generated by
    \textit{contbin}. Bottom: results for regions created in the subsequent rebinning
    process. The second temperature shown for region B1 (open circle) is the result 
    obtained when the embedded cool region B9 is included.  \label{figxtemp1}}
\end{figure*}

The results of this first-level binning step are shown in the top two panels of
Fig.~\ref{figxtemp1}. As expected, the relatively low SNR chosen leads to large
errors in the measured temperatures and to low $\chi^2$ values of $\sim
0.6$. Although the best-fit temperatures for some regions, mostly the hottest
ones, are poorly constrained, we already see clear variations and trends of the
ICM temperature within our study region.


\begin{deluxetable}{cccl}
\tabletypesize{\scriptsize}
\tablewidth{0pc}
\tablecolumns{4} 
\tablecaption{Results from spectral fitting \label{tablextemp}}
\tablehead{
\colhead{Region}    & \colhead{kT\tablenotemark{*}}         & \colhead{$\chi^{2}$} & \colhead{Binning Table}  \\
\colhead{}                     & \colhead{(keV)}&  \colhead{}                   & \colhead{}   
}
\startdata

     B1 &    $24.3\pm 5.3$  &    0.78 & A1, A5,  A7 \\
     B2 &    $16.1\pm 4.0$  &    0.57 & A2, A3,  A4, A8 \\
     B3 &    $14.8\pm 2.9$  &    0.56 & A6, A9, A15 \\
     B4 &    $12.9\pm 4.1$  &    0.69 & A10,  A11, A13 \\
     B5 &    $10.9\pm 1.7$  &    0.71 & A20, A27, A28 \\
     B6 &    $10.2\pm 2.5$  &    0.73 & A12, A17, A19 \\
     B7 &    $10.2\pm 2.4$  &    0.71 & A14, A18, A26 \\
     B8 &    $ 8.7\pm 1.2$   &    0.84 & A16, A23, A32 \\
     B9 &    $ 8.4\pm 3.6$   &    0.70 & A22 \\
     B10&    $ 8.2\pm 1.5$  &    0.69 &  A21, A24, A25 \\
     B11&    $ 7.3\pm 0.7$  &    0.73 & A29, A30, A31, A34 \\
     B12&    $ 5.7\pm 0.8$  &    0.93 & A33 

\enddata
\tablenotetext{*}{The quoted errors denote $1\sigma$ uncertainty.}

\end{deluxetable}

\subsection{Re-binning}\label{xray-temp-2}

As a second step, we examined the temperature map from Stage 1 and combined
adjacent regions with similar temperature in order to obtain a clearer view of 
temperature variations on larger scales, the goal being to reach at least 1200 net counts per bin by
combining three (occasionally four) of the initial regions. The results of this
rebinning process are shown in the bottom two panels of Fig.~\ref{figxtemp1};
the individual regions and their temperatures are listed in
Table~\ref{tablextemp}.

Note that two of the initial regions were excluded from the rebinning
and left unchanged, namely the ones coinciding with the main peaks in the X-ray
surface brightness distribution, A22 (B9) and A33 (B12). Both of these exhibit
markedly cooler temperatures than their surroundings, making them likely
candidates for cool cores of clusters still in the process of merging with the
main system. This interpretation is particularly plausible for the region around 
the global peak in the X-ray emission from MACSJ0717.5+3745, A33 (B12), 
which is found to contain the coolest gas anywhere within our study region 
(k$T=5.7\pm 0.8$ keV). By comparison, the temperature measured for the 
second putative cool core, region A22 (B9), is much higher and less well 
constrained (k$T=8.4\pm 3.6$ keV), but still in stark contrast to the much 
higher temperatures above 15 keV measured in its immediate vicinity
(regions A1, A5, and A7). With all of these regions being viewed in projection,
the true temperature difference is likely to be yet higher. Better constraints 
from a two-phase plasma model, especially relevant for region B9, would 
require significantly better photon statistics.

Although the selection of regions to merge is to some degree subjective, we
convinced ourselves that our final results and conclusions are robust. For
instance, our perhaps most debatable choice, the exclusion of the aforementioned
cool core A22 (B9) from the merging of the surrounding regions A1, A5, and A7
(B1), predictably increases the temperature measured for B1 (k$T=24.3\pm 5.3$
keV); including it, however, still leads to a statistically consistent result of
k$T=21.0\pm 5.4$ keV (Fig.~\ref{figxtemp1}).

\section{Cluster galaxy distribution and dynamics}\label{galprop}

\begin{figure}[h]
  \epsscale{1.0} \plotone{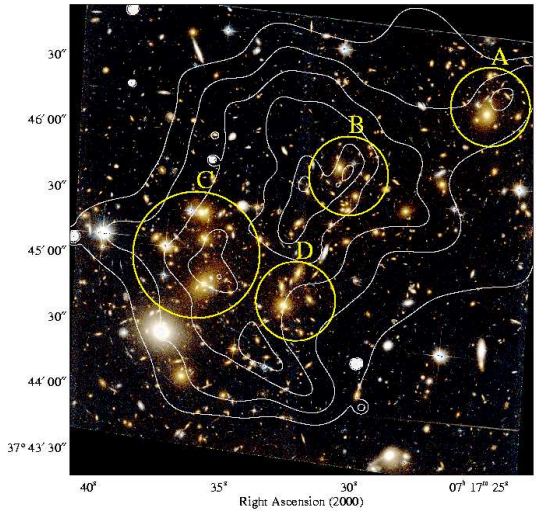} \figcaption[hstimage]{X-ray
    contours as in Fig.~\ref{figxtemp1} overlaid on the F555W/F814W optical
    image of the cluster core obtained with HST/ACS. Yellow circles centered on
    the four main peaks in the cluster light distribution mark the regions
    within which average radial velocities and velocity dispersions were computed
    for the enclosed galaxies (see Table~\ref{tablevel}).\label{fighst}} \vspace{-3mm}
\end{figure}

Our tentative interpretation of the physical origins of the complex temperature
and gas density distribution presented in the previous section can be tested 
and extended by studying the distribution and dynamics of the cluster galaxies 
in the same region. Shown in Fig.~\ref{fighst} is an overlay of the X-ray contours 
from Fig.~\ref{figxtemp1} onto the HST image of the cluster core. We identify four
major concentrations in the integrated optical light, which is dominated by the
large cluster ellipticals. Average redshifts and velocity dispersions for
galaxies within the four circles marked in Fig.~\ref{fighst} are listed in
Table~\ref{tablevel}. Unlike their counterparts in regions A, C, and D, the
galaxies associated with feature B are found to move through the cluster core at
a very high relative radial velocity of over 3,000 km s$^{-1}$, close to the
maximal velocity expected from infall from infinity. No significant differences
are found between the velocity dispersions measured for the four regions, 
although the results suggest that the structure in region C is the most massive one.

\begin{deluxetable}{cccc}
\tabletypesize{\scriptsize}
\tablewidth{0pc}
\tablecolumns{4} \tablecaption{Average radial velocities and velocity
  dispersions for galaxies in the regions marked in
  Fig.~\ref{fighst}.\label{tablevel}} \tablehead{
  \colhead{Label}           & \colhead{Num of redshifts}  & \colhead{$\Delta$V\tablenotemark{*} }  & \colhead{$\sigma$\tablenotemark{*} }    \\
  \colhead{} & \colhead{} & \colhead{(km/sec)} &\colhead{(km/sec)} } \startdata
A  &   10  &  $   278_{-339}^{+295} $ &  $ 1021_{-318}^{+139}$ \\
B  &    9   &  $  3238_{-242}^{+252} $ &  $  664_{-304}^{+ 63}$   \\
C  &   10  &  $  -733_{-478}^{+486} $ &  $ 1761_{-607}^{+234}$ \\
D  &    7   &  $   831_{-800}^{+843} $ &  $ 1328_{-424}^{+968}$
\enddata
\tablenotetext{*}{Calculated using the bootstrap estimator in \citet{beers90}.}
\end{deluxetable}

\section{Interpretation}

A comparison of the gas and galaxy distribution shown in Fig.~\ref{fighst} as 
well as of the radial velocity information summarized in Table~\ref{tablevel} 
yields important insights into the dynamics of the multiple
components of the cluster core. Specifically, we note that, of the four galaxy
concentrations discussed in Section~\ref{galprop}, only A and C are reasonably
well aligned with peaks in the X-ray surface brightness. The good X-ray/optical
alignment of component C, combined with a systemic velocity and velocity
dispersion consistent with those of MACSJ0717.5+3745 as a whole, makes this
subsystem a likely candidate for the actual, if disturbed, core of the main
cluster.

The dominant galaxies in regions B and D, on the other hand, are offset by about
25 arcsec ($\sim$160 kpc at the cluster redshift) from the respective X-ray peak
-- in both cases in a direction that points toward the filament to the SE of the
main cluster. Offsets of this kind between gas and galaxies (and dark matter)
have been observed before in merging clusters,  examples being the Bullet
Cluster 1ES0657--56 \citep{markevitch04} and MACSJ0025.4--1222
\citep{bradac08}, and are readily explained by the difference in the
self-interaction cross sections of the various cluster constituents. The
X-ray temperature maps of Fig.~\ref{figxtemp1} show the gas near regions B
and D being cool compared to their surroundings (B9 and B12); we can thus
safely identify them as the cool cores of clusters merging with the main body of
MACSJ0717.5+3745 and falling behind as their respective galaxy populations
advance. While the merger axis of system D is consistent with lying in
the plane of the sky (and aligned with the large-scale filament prominently
visible on the left of Fig.~\ref{xrayim}), the combination of the observed
X-ray/optical offset with a very high radial velocity indicates that system B is 
moving through the cluster core along an axis that is much more inclined 
toward our line of sight. We note also that the high-velocity collision of 
component B most likely happened only recently, given that its cool core 
(region B12 in Fig.~\ref{figxtemp1}) appears to be still largely intact.

A similar argument can be made to explain the observational evidence in region
A. The X-ray/optical offset suggests infall from the NW, i.e.\ in the direction
of the filament. The trailing peak in the gas density is highly elongated
though, and no cool gas is detected in the area, all of which suggests that we
are witnessing back-infall of a cluster that originated from the filament and
has already passed through the main cluster once.

While the combined X-ray/optical evidence allows a straightforward and
self-consistent interpretation of the dynamics and three-dimensional geometry of
the three seperate ongoing merger events described above, the gas distribution
and temperature structure near the cluster-filament interface -- the south-eastern 
corner of Figs.~\ref{figxtemp1} and \ref{fighst} -- defies simple explanation. 
Although the extremely high temperatures measured in regions B1 and B2 could to
some extent be caused by shock heating from the ongoing mergers of components B
and D, this explanation appears implausible for the entirety of the hot
interface region extending from B3 to B4. The sheer size of this feature (about
1 Mpc), its alignment perpendicular to the cluster-filament axis, and the
simultaneous increase in gas temperature and X-ray surface brightness all point 
to a direct large-scale interaction between the filament and the cluster core. We 
therefore speculate that the prominent jump in gas density and temperature along
the cluster-filament interface is the result of contiguous accretion of gas along the
large-scale filament feeding MACSJ0717.5+3745. A deeper Chandra observation
would allow this hypothesis to be tested and may yield the first credible 
characterization of the gas content of large-scale filaments.

\acknowledgements We thank Phil Marshall and the HST Archive Galaxy-scale
Gravitational Lens Survey (HAGGLeS) team for making their re-processed HST
images of MACSJ0717.5+3745 available for this study. We are also indebted to an
anonymous referee whose criticism led to substantial improvements of this
paper. Last, but not least, we gratefully acknowledge financial support by NASA
grant NAG 5-8253, SAO grant GO3-4168X, and STScI grant GO-09722.

\end{document}